\documentclass[a4paper,11pt]{article}
\pdfoutput=1 

\usepackage{jinstpub} 

\title{\boldmath Background identification algorithm for future self-triggered air-shower radio arrays}

\author[a,1]{S. Malakhov,\note{Corresponding author.}}
\author[a]{P. Bezyazeekov,}
\author[a]{O. Fedorov,}
\author[a]{Y. Kazarina,}
\author[b]{D. Kostunin,}
\author[c]{V. Lenok}

\affiliation[a]{Applied Physics Institute of ISU,Irkutsk, Russia}
\affiliation[b]{DESY, Zeuthen, Germany}
\affiliation[c]{Institute for Nuclear Physics, KIT, Karlsruhe, Germany}

\emailAdd{malakhov@astroparticle.online}

\abstract{
The study of the ultra-high energy cosmic rays, neutrinos and gamma rays is one of the most important challenges in astrophysics. The low fluxes of these particles do not allow one to detect them directly. The detection is performed by the measuring of the air-showers produced by the primary particles in the Earth's atmosphere. 
A radio detection of ultra-high energy air-showers is a cost-effective technique that provides a precise reconstruction of the parameters of primary particle and almost full duty cycle in comparison with other methods. 
The main challenge of the modern radio detectors is the development of efficient self-trigger technology,
resistant to high-level background and radio frequency interference. 
Most of the modern radio detectors receive trigger generated by either particle or optical detectors.
The development of the self trigger for the radio detector will significantly simplify the operation of existing instruments and allow one to access the main advantages of the radio method as well as open the way to the construction of the next generation of large-scale radio detectors.
In the present work we discuss our progress in the solution of this problem, particularly the classification of broadband pulses.}


\proceeding{International Conference on Instrumentation for Colliding Beam Physics\\
  Budker Institute of Nuclear Physics, Novosibirsk, Russia\\
 24 February~--- 28 February 2020}

\begin{document}
\maketitle
\flushbottom

\section{Introduction}
\label{sec:intro}

Radio detection of cosmic-ray air-showers is the promising technique providing cost-effective and high duty-cycle measurements.
Although radio emission from extensive air-showers (EAS) has been detected and studied since a long time, the technique becomes applicable for the field measurements only from the beginning of this century with the progress of modern digital data acquisition and processing systems.
Most ground-based EAS radio arrays perform measurements jointly with master detectors (particle or optical detectors), which produce the trigger for the entire facility.
The main challenge for the development of self-triggered radio array is a low signal-to-noise ratio (SNR).
Plenty of radio frequency interference (RFI), natural and man-made background distort and overlap EAS signal.
Impossibility of independent radio measurements leads to the necessity of supporting each radio array with particle or light detectors, and consequently to the impossibility of building large cost-effective radio arrays.
At the present moment there is no successful operating self-triggered ground-based radio EAS array, except ARIANNA~\cite{Barwick:2016mxm}, which indeed operates in extremely radio-quiet areas.
Besides it, research in this direction is carried out at few experiments, namely OVRO-LWA~\cite{ovro} and LOFAR~\cite{lofar}.

The main bottleneck in the radio detection of EAS is the increased flow of raw data comparing to other detectors, since the analysis of radio data requires recording of the full electrical field from each antenna station (cf. charges or waveforms from particle and Cherenkov detectors).
Self trigger for radio requires real-time processing of raw data with the rate of about GB/s for a single antenna station. To decrease the resulting data flow dumped from the fast ADC it is also necessary to apply additional online reconstruction (e.g. arrival direction).
The proposed pipeline can be implemented using FPGA technology.
In present work we discuss the architecture of radio trigger and algorithms for classification of broadband pulses.

\section{Tunka-Rex}
\label{sec:trex}
Tunka Radio Extension (Tunka-Rex)~\cite{trex} is a digital antenna array located in Eastern Siberia which measures EAS radio emission in the energy range of primary particle of $10^{17}$ - $10^{18}$ eV and frequency band 30-80 MHz.
Measurements are triggered by the host-detectors Tunka-133 (air-Cherenkov array) and Tunka-Grande (scintillator array) of TAIGA~\cite{Kostunin:2019nzy}, depending on the operation mode.
Array consists of 63 antennas covering an area of about 3 km$^{2}$.
Antenna station is based on SALLA (Short Aperiodic Loaded Loop Antenna)~\cite{salla} measures two perpendicular horizontal polarizations of EAS pulse.
The full signal chain has the following structure: SALLA $\to$ 30 meters cable $\to$ analog filter-amplifier $\to$ ADC with ring buffer (200 MHz 12 bit) $\to$ DAQ.
When the trigger is received, Tunka-Rex records the radio traces with a duration of 5 {\textmu}s from ring buffer for each station and sends them to the DAQ.
Signal of EAS has a duration of few tens of nanoseconds and located in a specific region in the recorded trace, namely "signal window".
Timestamp of signal in the recorded trace for each station is different, but known by cable lengths and hardware delays.
Standard Tunka-Rex data processing and reconstruction pipeline of EAS parameters is performed using timestamp and arrival direction given by the host detector.
For a detailed explanation of recent quality cuts and reconstruction procedure see Ref.~\cite{trex-template}.
During last years all data collected by Tunka-Rex are stored in Tunka-Rex Virtual Observatory (TRVO)~\cite{trvo}, a database with fast access to different data layers. Using this database we develop and test methods for independent, self-triggered detection of EAS pulses in raw data flow.

\section{Architecture of self trigger}
\label{sec:arch}
The first level of independent EAS pulse detection is a threshold trigger on channel-, station- and cluster-level.
In ideal radio-quiet conditions it is possible to perform measurements using this method only, but in a realistic environment we have to apply additional processing before triggering and recording, because a significant amount of transient RFI in conjunction with limited bandwidth of DAQ communication lines and storage capacity leads to an incredible fraction of dead time. 

The bottleneck of modern data acquisition systems is data transmission channels to the data center.
The use of a trigger on a threshold leads to the frequent trigger generation with a low fraction of true positives, which will cause the overloading of the data acquisition system.
Of course, this problem can be solved using fast network equipment, powerful servers and large storage, but this led to a significant increase in the costs of the final setup.
To reduce the load on the transmission channels it is necessary to implement hardware rejection of background data.
This step includes tasks of developing methods for deciphering the nature of the short radio pulses for separating the noise from EAS pulse as well as further hardware implementation of this method.

In the classical approach for radio detection we use advanced offline analysis, but this analysis does not applicable for real-time trigger generation.
For hardware trigger implementation we need a set of simple algorithms that can be easily implemented on FPGA.

Our approach suggests using compact antenna clusters and multi-level trigger generation scheme.
Experience has shown that using compact antenna clusters brings better fault tolerance and shows improved reconstruction accuracy.
The main idea of this approach is a gradual decrease in the data flow with a simultaneous complication of algorithms for searching of EAS signal.


\pagebreak

We suggest the following hierarchical structure of algorithms for real-time processing for the reduction of data flow and decreasing the dead time caused by readout:
\begin{enumerate}
     
\item \textit{Channel level.}
Improving signal quality using digital filters such as bandpass filter and median filter.
\item \textit{Station level.}
Two-channel analysis and data pre-selection.
The analysis includes SNR cut and RFI suppression algorithms.
RFI from a known source (usual case is other detectors located at the Tunka facility associated with pulses from the near-horizon area) can be easily suppressed by a specially tuned algorithm.
Trigger will be produced with the following conditions: amplitude and SNR ratio of the signal exceed the given thresholds and pulse is not associated with known RFI sources.
\item \textit{Cluster level.}
Pre-selection of station-level data by cluster-level analysis.
The analysis includes matched filtering, arrival direction cuts and suppression of known RFI by waveform. 
\end{enumerate}
\section{Methods of classification of broadband pulses}
\label{sec:methods}

Effective noise reduction is not possible without a clear understanding of the noise environment in the experiment.
The short pulses like EAS ones are the most difficult for filtering.
Such pulses can be generated by various kinds of pulsing electronics and they have stable form and amplitude.
To identify such pulses we developed a method for searching pulses from stable RFI sources.
The main idea of this method is an applying of filters tuned for RFI detection and analysis of amplitude and arrival direction of RFI.
On the first step we developed single station analysis (2 channels) and test it on Tunka-Rex data.
Like RFI detector we used root mean square (RMS) in the window of 640~ns.

\begin{figure}[t]
	\includegraphics[width=1\linewidth]{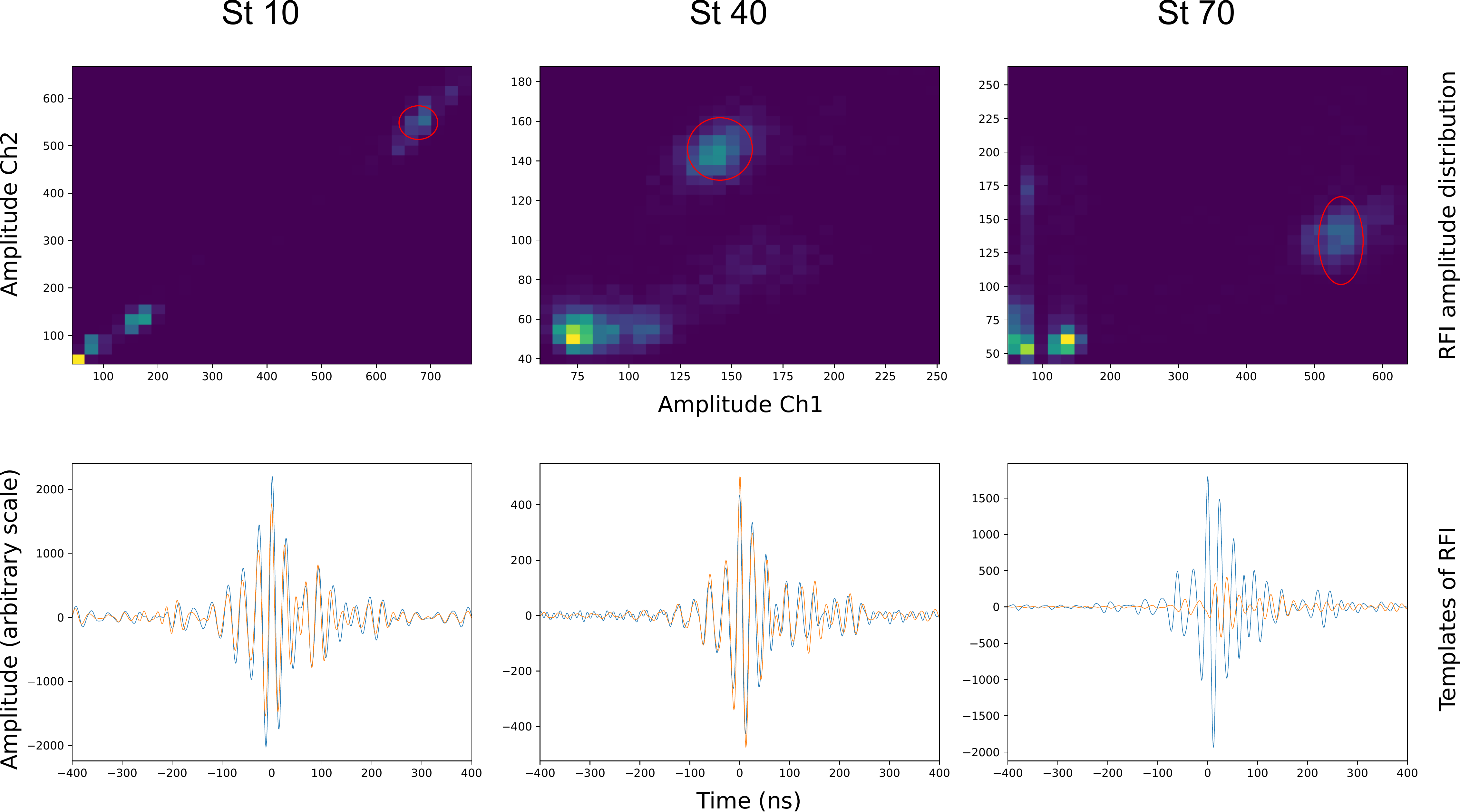}
	\caption{
	    \textit{Top:} examples of channel RMS distributions from neighboring Tunka-Rex stations.
	    \textit{Bottom:} Pulses obtained by averaging of signatures indicated by red circles. 
	    One can see, that detected RFI are similar for all three antennas what can point to the location of the noise source.
	}
	\label{fig:RMSDistr}
\end{figure}

The method is defined as follows:
\begin{enumerate}
    \item Applying a sliding window for each recorded trace and calculation of RMS in this window separately for each channel.
    Window containing a maximum of RMS is defined as a window contains a candidate of RFI pulse.
    RFI position is defined as a position of the peak of Hilbert envelope in the corresponding window.
    \item Creating a channel-to-channel RMS distribution of RFI candidates for all processed data.
    \item Finding RFI cores in this distribution by iterative Gauss fit (see Fig.~\ref{fig:RMSDistr}).
    \item Averaging of signals inside each core of RMS distribution for getting the template of RFI pulse.
\end{enumerate}


Applying this method to the data recorded at a given station allows one to get specific types of high-amplitude pulses detected at this station.
After that high amplitude pulses are pooled to the pulse library specific for each station.
Pulse library will be used for rejection of high amplitude pulses that are not associated with EAS pulse.
Proposed criterion of rejection is the convolution between noise pulse and raw time trace surpassing the threshold.

\section{Discussion and conclusion}
\label{sec:concl}
We developed the method for detection and shape definition of RFI from stable sources by using data from only one antenna station.
Test using Tunka-Rex archival data (TRVO) showed the efficiency of this technique.
Our method can be easily adapted for various methods of RFI detection such as matching filtering with known RFI template and neural network processing.
Also, this technique can be adapted to search for EAS signals.

In continuation of this analysis we will develop methods for multi-station analysis for RFI identification.
Investigating the timing profile and lateral distribution function of RFI we can define the position of this noise source.
Moreover, analyzing the data from several antenna stations we can reconstruct the shape of RFI pulses in more detail.

The next steps on the way of independent radio triggering are optimization of algorithms of defining RFI pulses, making a library of specific pulses for each station, developing methods for filtering and testing its efficiency with TRVO data.
By this we will reduce raw data flow and be ready for developing and testing hardware prototype.

In addition to self-trigger technique, the algorithms developed in the frame of this work can complement the methods for the lowering the threshold, e.g. ones using deep learning~\cite{Bezyazeekov:2019jbe}, as well as for the future low-threshold detectors~\cite{Schroder:2018dvb} or precise 21cm cosmology operating in 30-80~MHz domain~\cite{Kostunin:2019gho}.

\acknowledgments

This work was  supported by the Russian Federation Ministry of Science and High Education (agreement \textnumero~075-15-2019-1631, project. FZZE-2020-0024), by the Russian Science Foundation Grant No. 19-72-00010 (section 2,3) and by Russian Foundation for Basic Research grant 18-32-20220.



\end{document}